\documentclass[twocolumn,tighten]{aastex63}

\usepackage{txfonts}
\usepackage{amssymb}
\usepackage{multirow}
\usepackage{float}
\usepackage{gensymb}
\usepackage{xcolor}
\usepackage{verbatim} 
\usepackage{rotating}

\received{January 27}
\revised{March 3}
\accepted{March 3}
\shorttitle{BHB2007-1}
\shortauthors{Zurlo et al.}

\begin{document}


\title{Near-IR observations of the young star [BHB2007]-1:\\ A sub-stellar companion opening the gap in the disk}

\correspondingauthor{Alice Zurlo}
\email{alice.zurlo.astro@gmail.com}

\author[0000-0002-5903-8316]{Alice Zurlo}
\affiliation{N\'ucleo de Astronom\'ia, Facultad de Ingenier\'ia y Ciencias, Universidad Diego Portales, Av. Ejercito 441, Santiago, Chile}
\affiliation{Escuela de Ingenier\'ia Industrial, Facultad de Ingenier\'ia y Ciencias, Universidad Diego Portales, Av. Ejercito 441, Santiago, Chile}

\author[0000-0002-4266-0643]{Antonio Garufi}
\affiliation{INAF, Osservatorio Astrofisico di Arcetri, Largo Enrico Fermi 5, I-50125 Firenze, Italy}


\author[0000-0003-2953-755X]{Sebasti\'an P\'erez}
\affiliation{Departamento de F\'isica, Universidad de Santiago de Chile. Avenida Ecuador 3493, Estaci\'on Central, Santiago, Chile}
\affiliation{Center for Interdisciplinary Research in Astrophysics and Space Exploration (CIRAS), Universidad de Santiago de Chile, Estaci\'on Central, Chile}

\author[0000-0002-7945-064X]{Felipe O. Alves}
\affiliation{Max-Planck-Institut f\"ur extraterrestrische Physik, Gie{\ss}enbachstra{\ss}e  1, Garching, 85748, Germany}


\author[0000-0002-3829-5591]{Josep M. Girart}
\affiliation{
Institut de Ci\`encies de l’Espai (ICE), CSIC, Can Magrans s/n, 08193 Cerdanyola del Vall\`es,  Catalonia, Spain}
\affiliation{Institut d’Estudis Espacials de Catalunya (IEEC), 08034 Barcelona, Catalonia, Spain}

\author[0000-0003-3616-6822]{Zhaohuan Zhu}
\affiliation{Department of Physics and Astronomy, University of Nevada, 4505 South Maryland Parkway, Las Vegas, NV 89154, USA}

\author[0000-0003-2020-2649]{Gabriel A. P. Franco}
\affiliation{Departamento de F\'isica–ICEx–UFMG, Caixa Postal 702, 30.123-970 Belo Horizonte, Brazil}

\author{L. Ilsedore Cleeves}
\affiliation{Department of Astronomy, University of Virginia, Charlottesville, VA 22904, USA}





\begin{abstract}
The presence of planets or sub-stellar objects still embedded in their native protoplanetary disks is indirectly suggested by disk sub-structures like gaps, cavities, and spirals. However, these companions are rarely detected. We present VLT/NACO high-contrast images in $J$, $H$, $K_S$, and $L^{\prime}$ band of the young star [BHB2007]-1 probing the inclined disk in scattered light and revealing the probable presence of a companion. The point source is detected in the $L^{\prime}$ band in spatial correspondence with complementary VLA observations. This object is constrained to have a mass in the range of { 37-47} M$_{Jup}$ and is located at 50 au from the central star, inside the 70 au-large disk cavity recently imaged by ALMA, that is absent from our NACO data (down to 20 au). This mass range is compatible with the upper end derived from the size of the ALMA cavity. The NIR disk brightness is highly asymmetric around the minor axis, with the southern side 5.5 times brighter than the northern side. The constant amount of asymmetry across all wavelengths suggests that it is due to a shadow cast by a misaligned inner disk. The massive companion that we detect could, in principle, explain the possible disk misalignment, as well as the different cavity sizes inferred by the NACO and ALMA observations. The confirmation and characterization of the companion is entrusted to future observations. 
\end{abstract}



\section{Introduction} \label{sec:intro}

Understanding planet formation is one of the main challenges of modern astronomy. Despite the rapidly increasing number of detected exoplanets in the last years, little is known about {\it where} and {\it when} planets form. In recent years, the high contrast imaging community has pushed to image planetary systems during their first few Myr. This technique provides a snapshot of an entire planetary system, and of the interaction between the new born planets and their protoplanetary disk \citep[see, e.g.,][]{2010Sci...329...57L, 2010Natur.468.1080M, 2018A&A...617A..44K}. { Lately, it was even possible to look for planets} in the process of accreting material, having a closer look on the formation process itself \citep{2018ApJ...863L...8W, 2019NatAs...3..749H, 2019A&A...622A.156C, 2020A&A...633A.119Z}. The two planets around the star PDS\,70 are the first strong evidence that planets may be responsible for gaps detected in protoplanetary disks, as they are found in the middle of the big cavity detected in scattered light \citep{2018A&A...617A..44K}. Apart from PDS\,70, many other disks present features that can be explained by the indirect influence of one or more planets, such as spirals, gaps, and cavities \citep[e.g.,][]{2018ApJ...869L..41A,2019ApJ...870...72D}, but direct detection (thermal emission or H$\alpha$) has been challenging \citep{2015A&A...573A.127C, 2020A&A...633A.119Z, 2020arXiv200706573V}.  

In this paper, we present VLT/NACO observations of two sources: 2MASS J17110392-2722551, also known as [BHB2007]-1 \citep{Covey2010} following the \citet{Brooke2007} naming convention and 2MASS J17110411-2722593 ([BHB2007]-2). Both stars are part of the Barnard 59 region, which is the  largest  and  most  massive dense  core  in  the  Pipe Nebula \citep{Forbrich2009}, located at a distance of 163 pc from us \citep{Dzib2018}. The new Gaia DR3 parallax of [BHB2007]-2 translates into a distance of 166 pc. [BHB2007]-1 is classified as a flat-spectrum object while [BHB2007]-2 is a Class II object \citep{Brooke2007}. \citet{Covey2010} inferred a spectral type of K7 for [BHB2007]-1 and M3 for [BHB2007]-2. [BHB2007]-1 is part of a small cluster of young stellar objects \citep{Brooke2007} in a low extinction region. The population of young stellar objects in the cluster includes young (Class 0/I) protostars such as [BHB2007]-11 \citep{Hara2013,Alves2017,Alves2019} and more evolved ones, like [BHB2007]-1 and [BHB2007]-2.

Recent observations by \citet{Alves2020} show a highly inclined (75 deg) disk around [BHB2007]-1. The ALMA continuum also shows a wide gap in the disk. Extended filaments are detected in molecular (CO) gas, which seem to feed the disk itself. VLA 22 GHz data report emission at the location of the primary star and a second fainter point source inside the dust gap. \citet{Alves2020} find that the gap size is consistent with being carved by a planet or a brown dwarf companion. To better investigate this scenario we present new NIR data of the system.

The analysis is presented as follows: in Sec.~\ref{Observations} we describe the observations and data reduction of the NIR data, in Sec.~\ref{res} our results on the analysis and interpretation of the data. In Sec.~\ref{Discussion} we present the discussion and we conclude in Sec.~\ref{conc} with a summary of the main results.

\section{Observations and data reduction} \label{Observations}
[BHB2007]-1 was first observed in the NIR with VLT/NACO during the night of 2004-04-30 under ESO program 073.C-0379(A) in the narrow band filter NB\_1.64, then again with the same instrument during the night 2008-07-01 under ESO program 081.C-0477(A). This second time the system was observed in filters $J$, $H$, $K_S$, and $L^{\prime}$. The observing conditions for the first epoch were not optimal, while during the second epoch the seeing was excellent, with a stable value of 0\farcs6. The exposure time was 46s in total for the first epoch in NB\_1.64, and for the second epoch 40s in $J$, and $\sim$5min each for $H$, $K_S$, and $L^{\prime}$. 

\begin{figure*}
\centering
\includegraphics[width=0.7\textwidth]{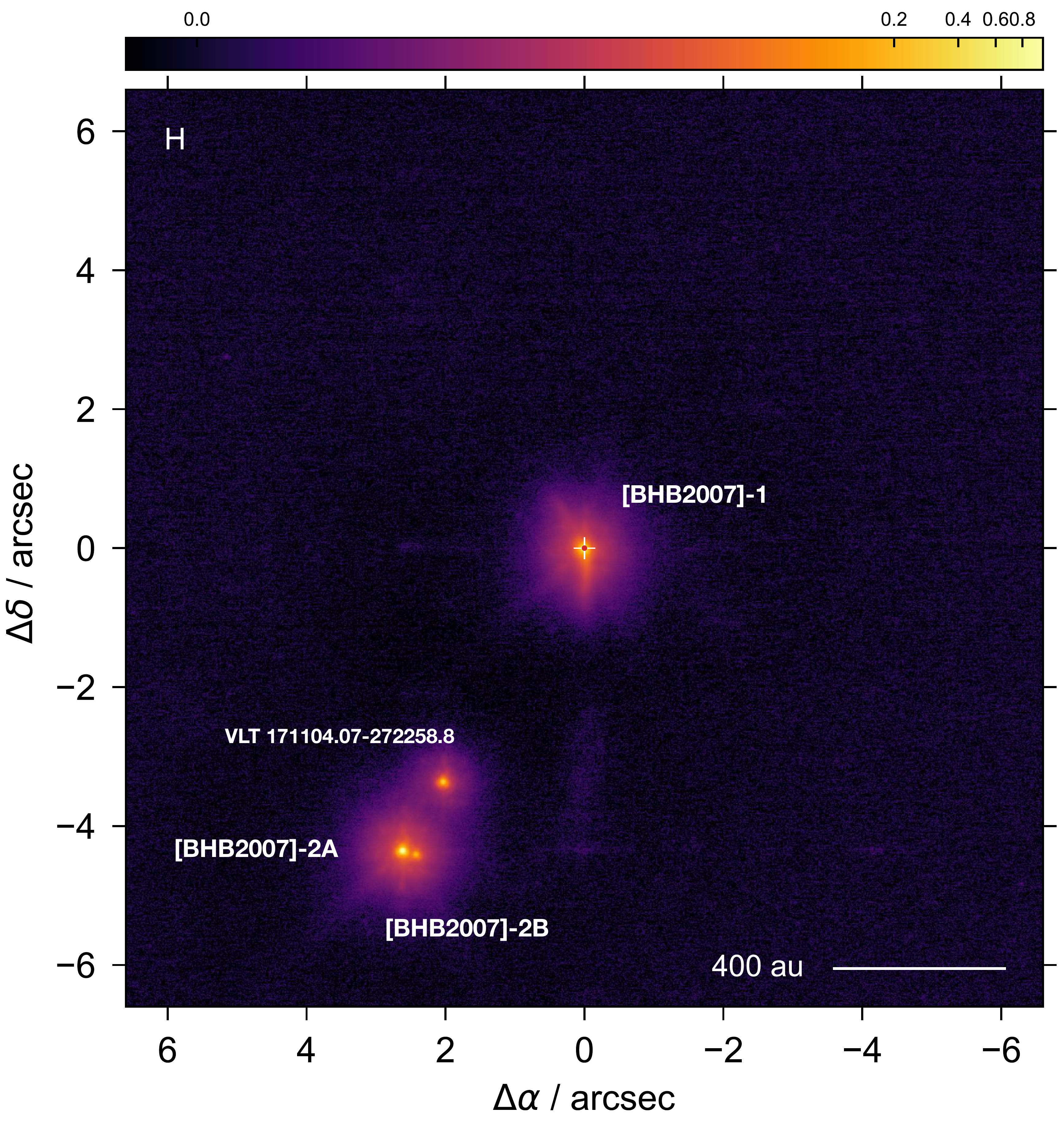}
\caption{Final NACO image in $H$ band. The whole detector is shown, where other objects are identified.} 
\label{f:naco_alls}
\end{figure*}

The observing strategy was the same for the two observing nights: the {\it jittering} technique, which consists of applying different offsets to have the star falling in different positions in the detector at each exposure. The data were reduced following \citet{2020MNRAS.496.5089Z} and \citet{2021MNRAS.501.2305Z}: the exposures were background subtracted (one from the median of the others), flat-fielded, and then recentered to have the target in the center of the detector. Each final image is the median of all the exposures combined together. During this procedure poor quality frames were excluded. For the astrometric calibration, we adopted the ALMA coordinates for the center of the star, RA=17:11:03.9234 and Dec=-27:22:55.4725, as measured in \citet{Alves2020}. We refer the reader to that publication for further details on the ALMA data on the source. The final $H$ band image where other 3 sources are visible in the detector is shown in Fig~\ref{f:naco_alls}. 

\begin{figure*}
\centering
\includegraphics[width=0.4\textwidth]{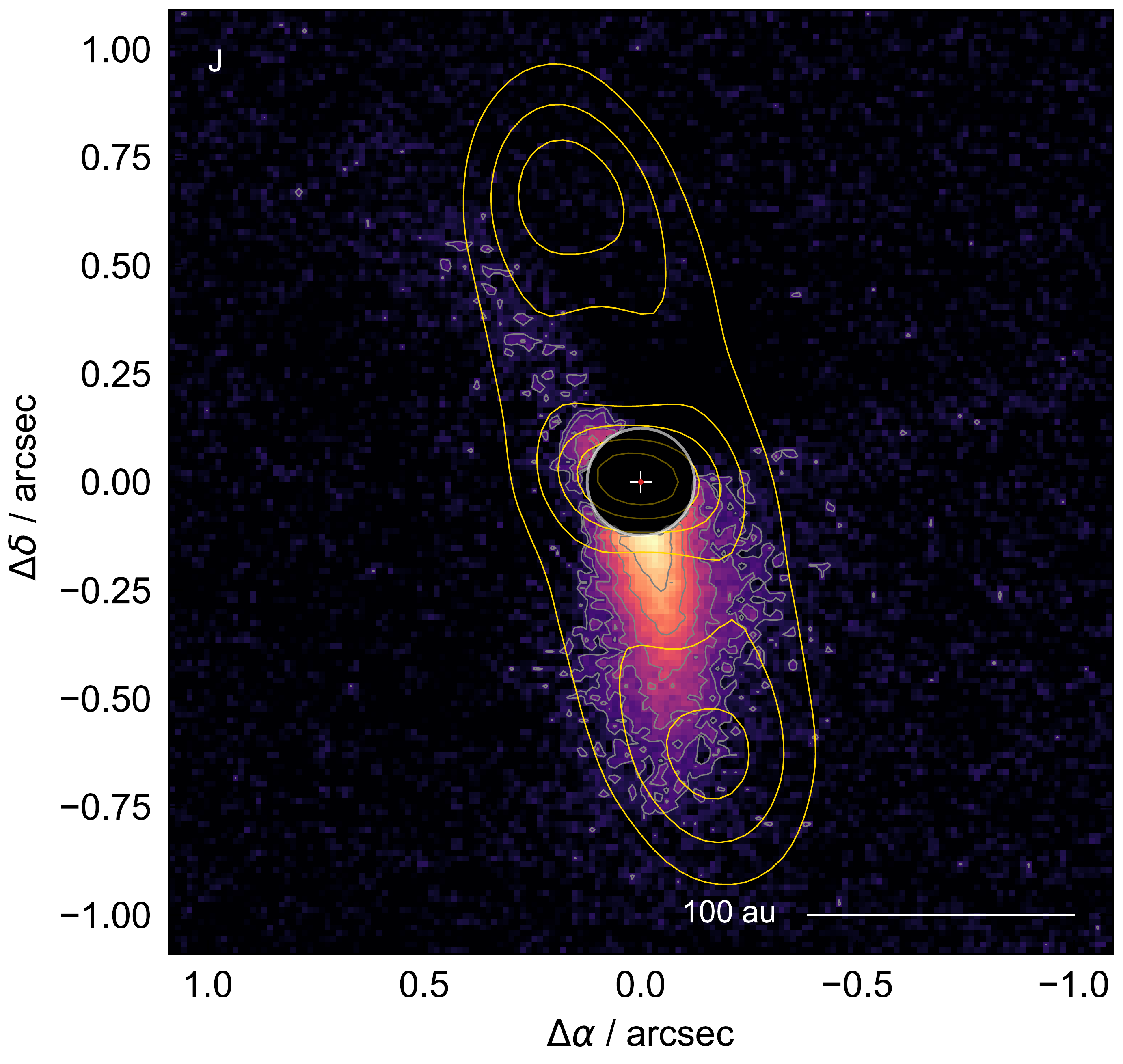}
\includegraphics[width=0.4\textwidth]{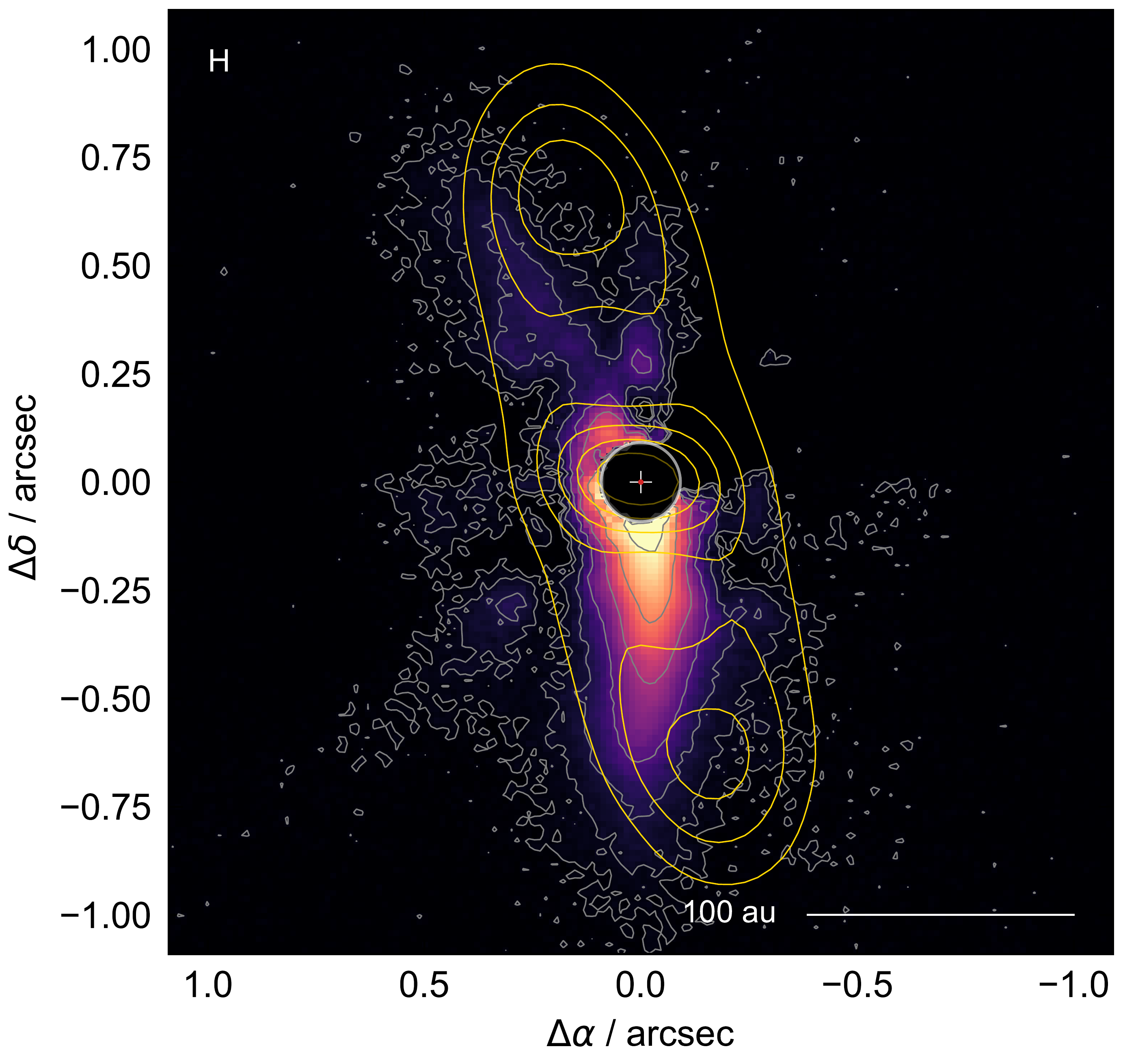}\\
\includegraphics[width=0.4\textwidth]{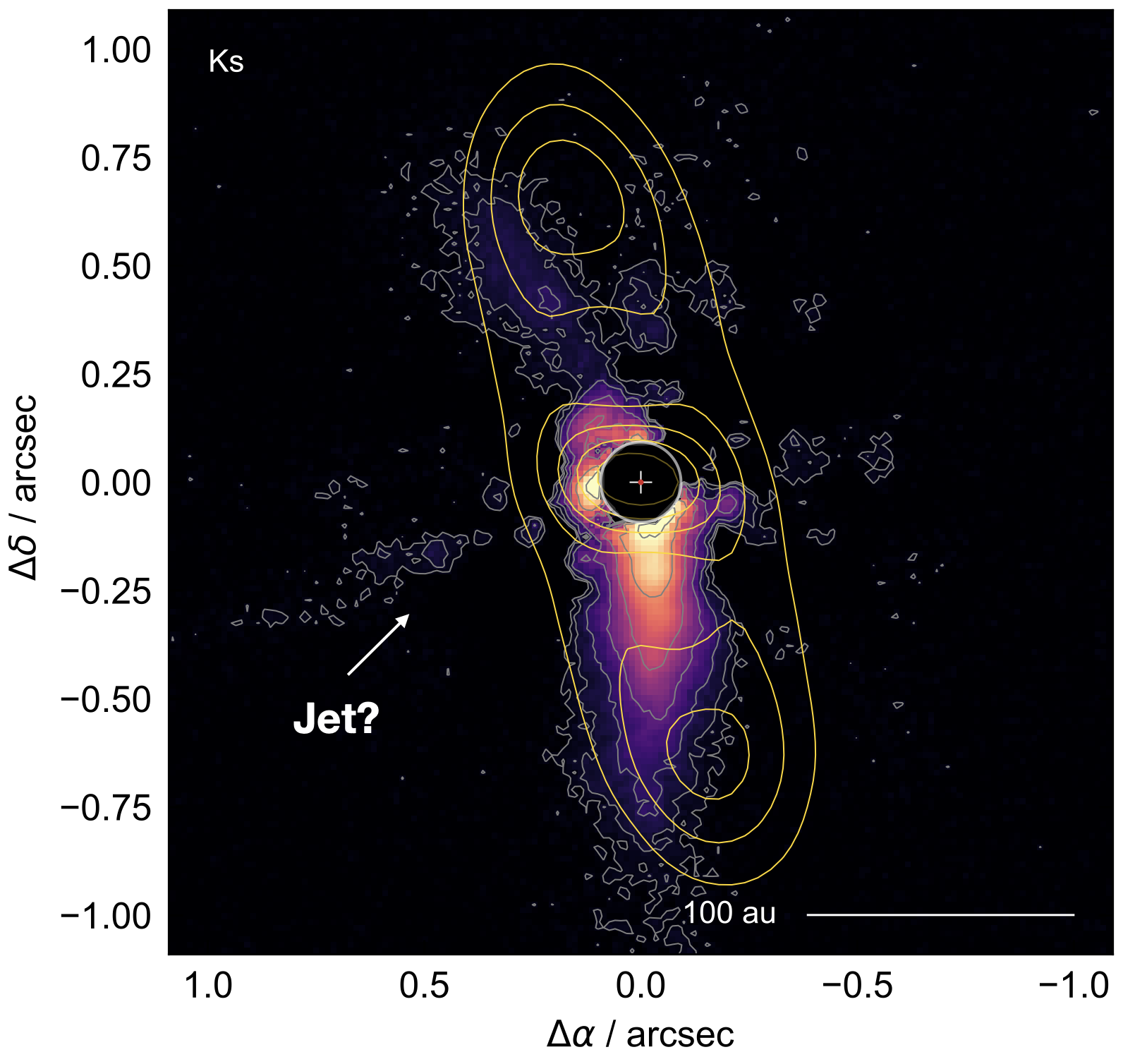}
\includegraphics[width=0.4\textwidth]{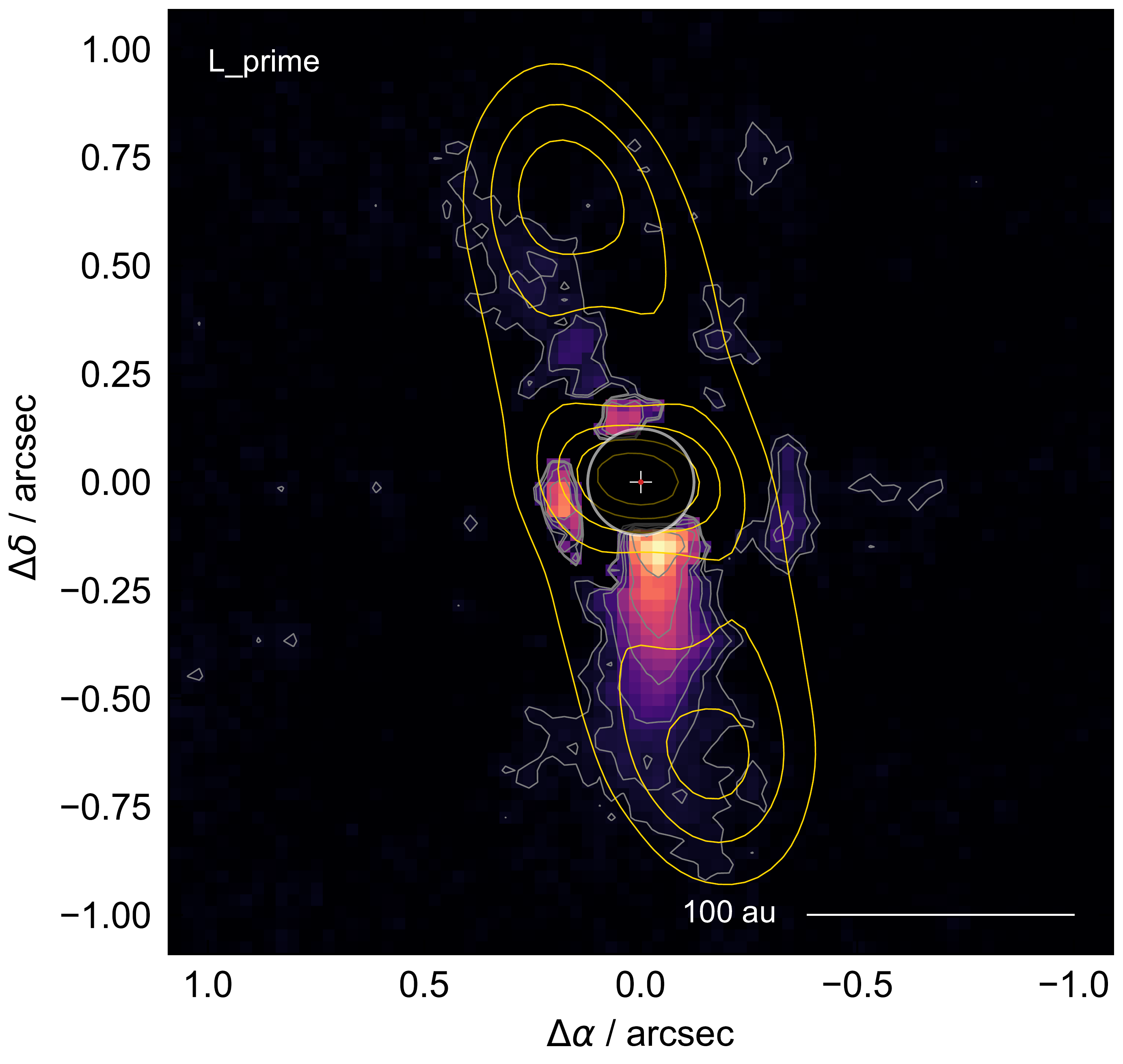}
\caption{An overlay between the NACO and ALMA observations of [BHB2007]-1. The yellow contours show the ALMA 1.3 mm continuum emission while the PSF subtracted NACO image is shown in false color with gray contours. The ALMA peak is matched to the NACO stellar position. { In the lower left panel, corresponding to the $K_S$ band, a possible jet feature is visible on the East of the primary, perpendicular to the disk.} }
\label{f:naco}
\end{figure*}

The observations, done without a coronagraph, show an unsaturated stellar point-spread-function (PSF).  In order to highlight extended emission around [BHB2007]-1, the subtraction of the bright stellar component is performed by reconstructing the PSF from the final images themselves. First, a 2D elliptical Gaussian is fit to the central region to determine the stellar centroid. A radial PSF profile is then constructed by performing an azimuthal average over angles devoid of extended emission (between PAs 230$^\circ$ and 350$^\circ$ eastwards from the North direction). Finally, the azimuthally-averaged radial profile is expanded in full azimuthal range to produce a 2D image of the PSF.  This procedure produced qualitatively better results than subtracting an 2D Gaussian or a scaled version of the PSF given by another star in the field. In $L^\prime$ band, the PSF subtraction allows to remove the Airy rings. The PSF-subtracted images from the second epoch only are shown in Fig.~\ref{f:naco}, cropped around [BHB2007]-1, together with the ALMA contours from \citet{Alves2020}. The final NB\_1.64 filter image of the first epoch is not shown as it has a lower quality and it was used in the following analysis only for the astrometric measurements.  

\begin{figure}
\centering
\includegraphics[width=\columnwidth]{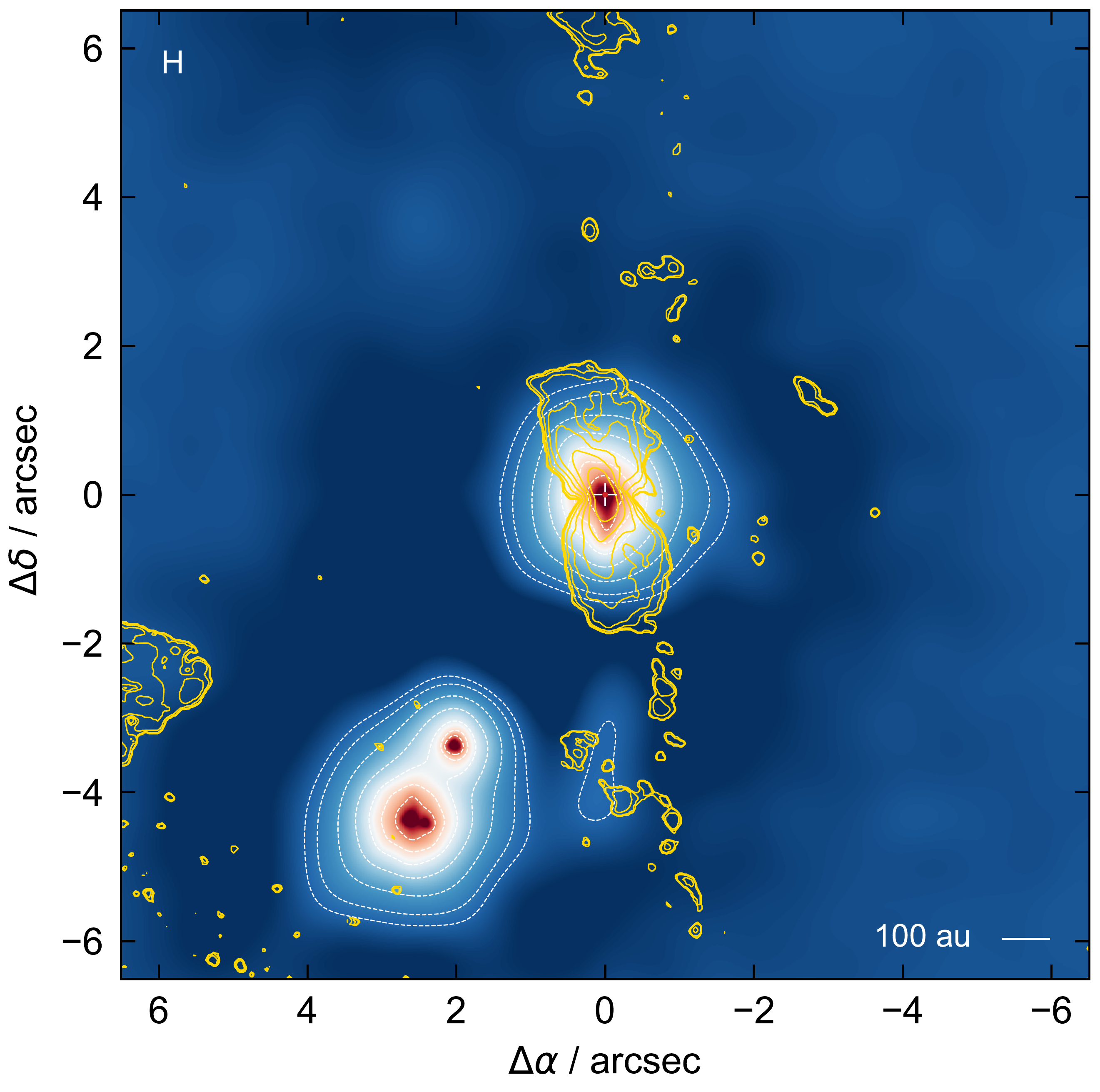}
\caption{An overlay between the $H$ band NACO and the CO gas emission ALMA observations of [BHB2007]-1. The yellow contours show the ALMA $^{12}$CO integrated emission. The NACO image has been processed using an adaptive kernel smoothing technique called {\sc denoise} \citep[part of the {\sc splash} suite,][]{2007PASA...24..159P} to highlight faint extended emission south of [BHB2007]-1  \citep[see][for an example of the usage of this denoising technique]{2020A&A...639L...1M}.  }
\label{f:naco_2}
\end{figure}

\section{Results}\label{res}

\subsection{Overview of the NACO images} \label{overview}

The NACO images reveal the existence of four stars in the field of view (see Fig.~\ref{f:naco_alls}). The astrometric and photometric properties of these four objects are analyzed in Sects.\,\ref{Astrometry} and \ref{Stellar_properties}, respectively. The closer object to the SE of [BHB2007]-1 appears for the first time in the early third Gaia release \citep[EDR3,][]{Gaia2020} and has been labeled as VLT 171104.07-272258.8 in Fig.~\ref{f:naco_alls}, following IAU standards. On the other hand, the known object [BHB2007]-2 is resolved here as a binary system, to our knowledge, for the first time. The less luminous star to the West of the binary system is named [BHB2007]-2B. The NIR images also reveal the existence of an inclined disk around [BHB2007]-1, of extended filaments, and of a possible sub-stellar object near [BHB2007]-1 that are analyzed in Sects.\,\ref{Disk} and \ref{Planet}. In the $K_S$ band a faint emission perpendicular to the disk plane is detected (see Fig.~\ref{f:naco}, lower left panel). If the feature is real, and not residual light from the spiders, it might be related with a jet. The jet may be an evidence that the protostar is actively accreting. Indeed, from the $JHK$ spectra presented in \citet{Covey2010}, both Br$\gamma$ and Pa$\beta$ lines, indicators of accretion, are seen in emission. { The NB\_1.64 image is centered at the [FeII] 1.64 $\mu$m line, which is a tracer of jets in young objects. However, the bad quality of our NACO observations does not allow to detect any emission from the potential jet.}

\subsection{Astrometry and photometry} \label{Astrometry}
In Gaia DR2 \citep{Gaia2018}, the only star in our full field of view (FoV) with astrometric measurements available was [BHB2007]-2, for which a parallax of $\pi$=5.9225 mas and a proper motions of $\mu_{RA}$=-1.411 mas/yr and $\mu_{Dec}$=-21.629 mas/yr were measured. In Gaia EDR3, the parallax measurement for the source [BHB2007]-1 is also given for the first time, $\pi$=4.30$\pm$0.52 mas. Proper motions are $\mu_{RA}$=-3.67 mas/yr and $\mu_{Dec}$=-17.94 mas/yr. For [BHB2007]-2 the parallax was updated to $\pi$=6.02$\pm$0.31 mas, $\mu_{RA}$=2.04 mas/yr and $\mu_{Dec}$=-21.95 mas/yr.

The EDR3 parallax of [BHB2007]-1 (translating into a distance of 232.6 pc) would place the source far behind the Pipe nebulae \citep[at 163 pc,][]{Dzib2018}. However, a few cases of erroneous parallax measurements from Gaia have been reported for young stars \citep[e.g.,][]{2019A&A...624A...4M, Garufi2019}. The Gaia astrometric excess noise is relatively large ($\sim$4 mas). { We checked for the fidelity of the EDR3 astrometric solutions for BHB2007-1 (sources ID 4108624199978985984) and BHB2007-2 (source ID 4108624195634062464) by querying the neural network classifier recently made available by \citet{2021arXiv210111641R}.  The astrometric fidelities are 0.9995 and 0.05053 for BHB2007-2 and BHB2007-1, respectively, meaning that the astrometric solution of the latter source is most likely spurious.} Furthermore, the systemic velocity inferred for [BHB2007]-1 by \citet{Alves2020} from CO millimeter lines (3.6 km s$^{-1}$) matches the systemic velocity of the other members of the Barnard 59 region. At 232 pc the disk and gap sizes resolved in the ALMA dust maps would be considerably larger than typical values for protoplanetary disks \citep{2019ApJ...872..112V}. Therefore, we are inclined to consider the Gaia EDR3 parallax erroneous and assume a distance of 166 pc, like [BHB2007]-2.

To assess whether VLT 171104.07-272258.8 (for which no Gaia parallax is available) is physically bound to the system, we exploited the NACO images from two different epochs ($\sim$4 years apart). For both images, we measured the accurate position of the three systems: the central star [BHB2007]-1, VLT 171104.07-272258.8, and [BHB2007]-2 AB, treated for this calculation as a bound binary system for the very close separation of the two components. 
The positions and relative motion of VLT 171104.07-272258.8, { [BHB2007]-2 A, and [BHB2007]-2 B} with respect to [BHB2007]-1, which is in the center of the detector in the two images, are listed in Table~\ref{t:astro}. Given the differences in RA and Dec we can say that VLT 171104.07-272258.8 is not co-moving with either of the two other systems visible in the detector. If bound, assuming a circular orbit, VLT 171104.07-272258.8 would orbit around [BHB2007]-2 on a period of $\sim$ 1700 yr, so we can easily exclude that the difference in RA and Dec on a 4 yr baseline are due to orbital motion. { On the other hand, [BHB2007]-2 A and B are a physically bound system as they show the same differences with respect to [BHB2007]-1, inside the errorbar. Note that one of the images used for this measurement is the narrow band filter NB\_1.64, which has a very poor quality, and [BHB2007]-2 B has a low SNR. { It is then difficult to infer any orbital motion between [BHB2007]-2 A
and B, due to the great errorbars. } This binary is not co-moving to { [BHB2007]-1}, they may have different proper motions ($\sim$10 mas/yr) if they are located at similar distance from us.} The ALMA continuum and molecular maps (CO, C$^{18}$O and H$_2$CO isotopologues) also do not show another source within the NACO FoV, indicating that both BHB2007-2AB and VLT 171104.07-272258.8 are more evolved than BHB2007-1. In the analysis we also calculated the photometry { in $K_S$ band of BHB2007-2AB and VLT 171104.07-272258.8, listed in Table~\ref{t:photo}}.
\\

\begin{table*}
\caption{{Astrometric measurements of the components measured with respect to the central source [BHB2007]-1.  }} 
\label{t:astro}
\centering
\begin{tabular}{lccccc}
\hline
\hline
Object  & Date  & Separation (mas) & PA (deg) & $\Delta$ RA (mas) & $\Delta$ Dec (mas) \\

\hline
VLT 171104.07-272258.8 &2004-04-30  &   3999  & -210 & & \\
   &2008-07-01  & 3920  & -211 & -36$\pm$ 5  & -113$\pm$ 5\\
BHB2007-2 A    &2004-04-30  &  5090 & -210 &  & \\
&2008-07-01   &  5070  & -211 & -42$\pm$ 5  &  -48$\pm$ 5 \\
{ BHB2007-2 B }   &2004-04-30  &  5039 & -208 &  & \\
 &2008-07-01   &  5024  & -209 & -46$\pm$ 7  &  -42$\pm$ 7 \\
\hline
\end{tabular}
\end{table*}

\begin{table}
\caption{{ Apparent magnitudes of the components measured with respect to the central source [BHB2007]-1.  }} 
\label{t:photo}
\centering
\begin{tabular}{lcccc}
\hline
\hline
Object  & $J$ & $H$   &   $K_S$ & $L^{\prime}$  \\

\hline
VLT 171104.07-272258.8 & 11.95  & 10.23 &  9.27 & 10.71 \\  
BHB2007-2 A   & 10.13 & 8.80 & 8.18 & 9.40\\
BHB2007-2 B  & 11.34 & 10.72 &  10.08  & 10.97 \\
\hline
\end{tabular}
\end{table}

\subsection{Stellar properties} \label{Stellar_properties}
Here we calculate the stellar properties of [BHB2007]-1 and -2 based on the Gaia EDR3 parallax (see Sec.\,\ref{Astrometry}) and the notion that [BHB2007]-2 is a binary system. We collected the B-to-W4 band photometry of -1 and -2 from \textit{Vizier}. We investigated the ASAS-SN light curves \citep{Kochanek2017} finding that the visual brightness of both stars is only moderately variable ($\lesssim$1 mag over years). We adopted a Phoenix model of the stellar photosphere \citep{Hauschildt1999} with the effective temperature constrained by \citet{Covey2010} ($T_{\rm eff}=4060$ K and $T_{\rm eff}=3346$ K, { for -1 and -2, respectively, with typical uncertainties of $\pm$300 K}) and visual extinction $A_{\rm V}$ calculated from the observed colors between $R$, $I$, and $J$ bands, and stellar luminosity $L_*$ from the integrated photospheric model. For [BHB2007]-1, this yielded $A_{\rm V}=1.9\pm1.0$ mag and $L_*=0.63\pm0.21\ \rm L_{\odot}$. {However, the edge-on geometry of the disk (see Sect.\,\ref{Disk}) may in principle yield an under-estimation of the luminosity. In fact, stars with very inclined disks appear on average older (since less luminous) than the other stars \citep[see][]{Garufi2018}.}

As for [BHB2007]-2, we first had to consider the visually triple nature of the star (see Sec.\,\ref{overview}). All three stars sit within the 2MASS beam, meaning that the observed photometry is a contribution of all of them. We computed the actual J, H, K$\rm _S$, and W1 band magnitude through the relative flux ratios measured in our NACO images. We then constrained visual extinction and stellar luminosity of [BHB2007]-2A assuming that the $T_{\rm eff}$ constrained by \citet{Covey2010} refers to this star since it is significantly brighter than the other two. We found that the $A_{\rm V}$ is comparable with that of [BHB2007]-1 ($1.5\pm0.4$ mag) while the luminosity is $0.58\pm0.16\ L_{\odot}$.

Finally, we constrained stellar mass $M_*$ and age $t$ of the two sources through multiple sets of pre-main-sequence tracks \citep[Baraffe, MIST, and Parsec,][]{Baraffe2015, Dotter2016, Bressan2012}. Bearing in mind the assumptions and the uncertainties described above, we found $M_*=0.65-0.85\ \rm M_{\odot}$ and $t=0.9-5.7$
Myr for [BHB2007]-1, and $M_*=0.24-0.30\ M_{\odot}$ and $t=0.1-0.8$ Myr for [BHB2007]-2. The stellar mass of [BHB2007]-1 is significantly smaller than what was constrained by \citet{Alves2020} from the position-velocity diagram of the CO emission (2.23 $\rm M_{\odot}$). 

Such a large discrepancy cannot be explained by an erroneous estimate of the stellar luminosity {alone} since the evolutionary tracks in correspondence of stars with $T_{\rm eff}=4000$ K are nearly vertical, making the stellar mass only marginally impacted by the luminosity. {A miscalculation of the effective temperature is also an unlikely explanation. In fact, a $T_{\rm eff}>5000$ K would be necessary to reconcile photometric and dynamical mass. \citet{Covey2010} obtained their estimate of 4060 K from near-IR spectral indices of good quality spectra making such a mismatch very implausible. A possible, and yet speculative, solution could be that the central star is in reality an unresolved binary with nearly equal-mass components \citep[see e.g., AK Sco,][]{Andersen1989, Czekala2015}. In such a configuration, the observed flux would be distributed to the two objects. Each component would have the same mass constrained for a single source because stellar tracks are vertical. The system would be calculated to be older than 5 Myr but the aforementioned underestimation of the stellar luminosity due to the disk inclination could place the system back to the young age of the Pipe Nebula.}

\subsection{Disk} 
\label{Disk}
In Figure~\ref{f:naco}, the light scattered off by the disk around [BHB2007]-1 is visible in spatial correspondence with the mm continuum emission presented by \citet{Alves2020}. The scattered light is detected in each of the four wavebands and is always stronger on the south side. To obtain a meaningful measurement of the disk brightness, we computed the disk-to-stellar light contrast $\phi=F_{\rm disk}\times 4\pi r^2 / F_*$, where $F_{\rm disk}$ is the flux measured from the PSF-subtracted image (see Sec.\,\ref{Observations}), $r$ is the distance from the central star, and $F_*$ is the stellar flux measured from the innermost resolution element of the star in the original intensity images \citep[see Appendix B of][for details]{Garufi2017}. This procedure reveals that the disk brightness decreases with increasing the wavelength (spanning on the bright side from 17.2\% to 6.0\%, see Table~\ref{Disk_brightness}). Nonetheless, the brightness ratio between the southern and northern sides is constant (with an average ratio of 5.6). The implications of this finding are discussed in Sec.\,\ref{Discussion}. 

\begin{table}
\caption{Disk brightness relative to the stellar brightness. For the disk north and south sides, the maximum disk-to-stellar light contrast (see Sec.\,\ref{Disk}) is shown. Errors are obtained from the standard deviation around the brightest pixel.} 
\label{Disk_brightness}
\centering
\begin{tabular}{lccc}
\hline
\hline
Waveband & South (\%) & North (\%) & Ratio \\
\hline
$J$ & 17.2$\pm$1.6  &  3.2$\pm$0.3  & 5.3$^{+1.2}_{-0.8}$ \\
$H$ & 12.4$\pm$1.2  &  2.1$\pm$0.2  & 5.9$^{+1.1}_{-1.0}$ \\
$K_{S}$ & 9.3$\pm$0.9  &  1.6$\pm$0.2  & 5.8$^{+1.5}_{-1.1}$ \\
$L^\prime$ & 6.0$\pm$0.7  &  1.1$\pm$0.1  & 5.4$^{+1.3}_{-1.0}$ \\
\hline
\end{tabular}
\end{table}

Beside its brightness, the disk looks similar across all images. It appears with a near edge-on inclination and is evidently flared. The outermost signal is detected at $\sim$ 1.0\arcsec, which is comparable with the disk extent constrained from the millimeter continuum emission by \citet{Alves2020}. On the other hand, our images do not reveal any disk cavity down to the NACO inner working angle of 0.12\arcsec\ ($\sim$20 au), in contrast with the millimeter cavity of $\sim$70 au inferred by \citet[see Fig.\,\ref{f:naco}]{Alves2020}. This discrepancy is routinely observed in protoplanetary disks and is further discussed in Sec.\,\ref{Discussion}.

\begin{figure}
\includegraphics[width=\columnwidth]{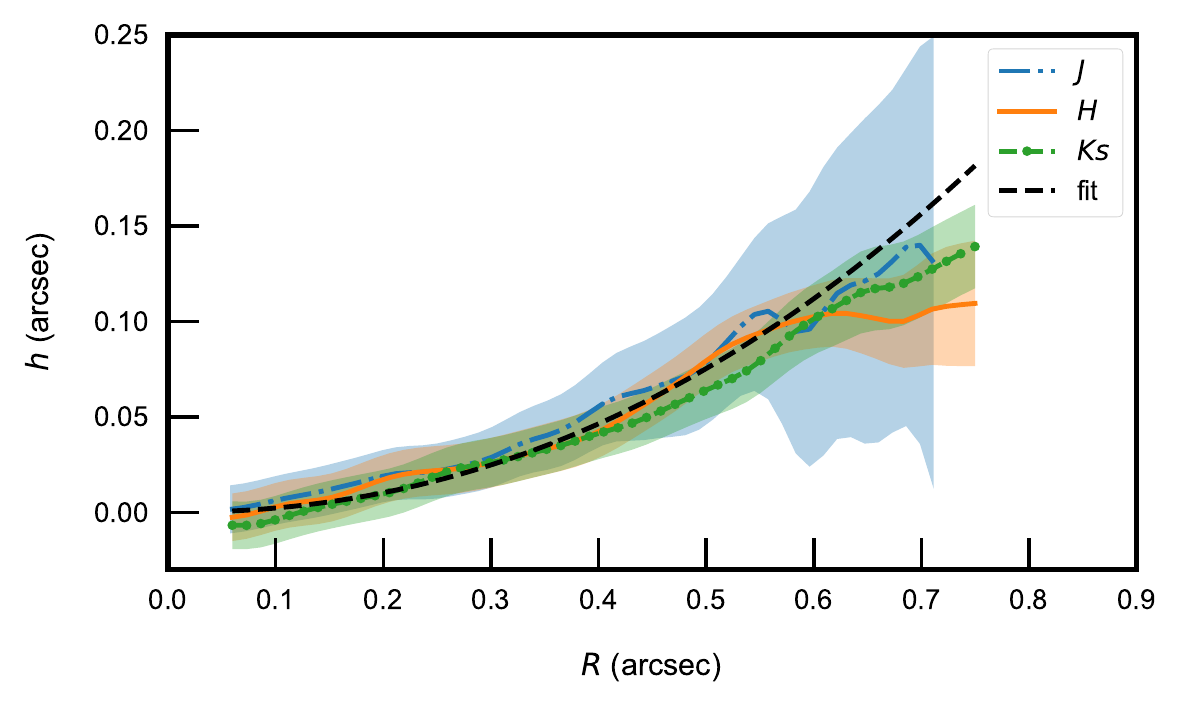}
\caption{Measurement of the height of the scattering surface as a function of distance from the central star. A power-law fit to the $H$-band profile yields a power-law index of $\sim$1.2 and $h(R=1\arcsec)=0.3$ (dashed line). The shaded areas correspond to 3$\sigma$ uncertainties around the measured profiles. These uncertainties are calculated as the quadratic sum of three terms: the error in the centroid fit, the width of the fitted Gaussian divided by the local signal-to-noise ratio, and a third term where we assume that the peak locations cannot be known to a precision better than a 10th of the image resolution. }
\label{f:heights}
\end{figure}

Since the disk is close to edge-on, we are able to measure the height of the scattering surface as a function of radius, for each observed wavelength. We do this by calculating the distance between the peak of the disk signal and the disk midplane. Here, we assume that the midplane is given by the disk PA of 12.5$^\circ$, as determined from the ALMA continuum by \citet{Alves2020}. We trace the disk surface by fitting a Gaussian as a function of height, for each pixel along the disk PA. We only considered the brighter (south) side of the disk in the calculation. The height profile is then deprojected by the disk inclination by dividing by $\sin(i)$, where $i=75^\circ$ as reported in \citet{Alves2020}. The extracted heights as a function of radius are shown in Fig.~\,\ref{f:heights}. The height profiles can be well approximated by a power-law in radius $h(R) \propto R^{1+\alpha}$. Assuming that the height of the scattering surface can be taken as an approximation to the vertical scale-height of the disk, the $\alpha$ index would be similar to the disk flaring. A simple fit to the $H$-band profile yields an $\alpha$ value of $\sim$1.2. The $\alpha$ value is lower ($\alpha\sim1$) for $J$ and $Ks$.

In the South of the object $\sim 4$\arcsec~below the NACO brightness peak a faint filament is seen in the $H$ band filter { with higher SNR with respect to the other filters}, which can be compatible to the signal seen in the $^{12}$CO by ALMA (see Fig.~\ref{f:naco_2} with the overlap of the two images). The confirmation that the filament seen in the NIR is not a NACO artifact came with the Gaia DR3, where signal is detected at the same position with a magnitude of $G$=20.77. The filament may also connect the system with [BHB2007]-2, which share the same distance. Indeed the filament seems to connect the two young protostars. 
\\


\subsection{Embedded sub-stellar object} \label{Planet}

\begin{figure}
\centering
\includegraphics[width=\columnwidth]{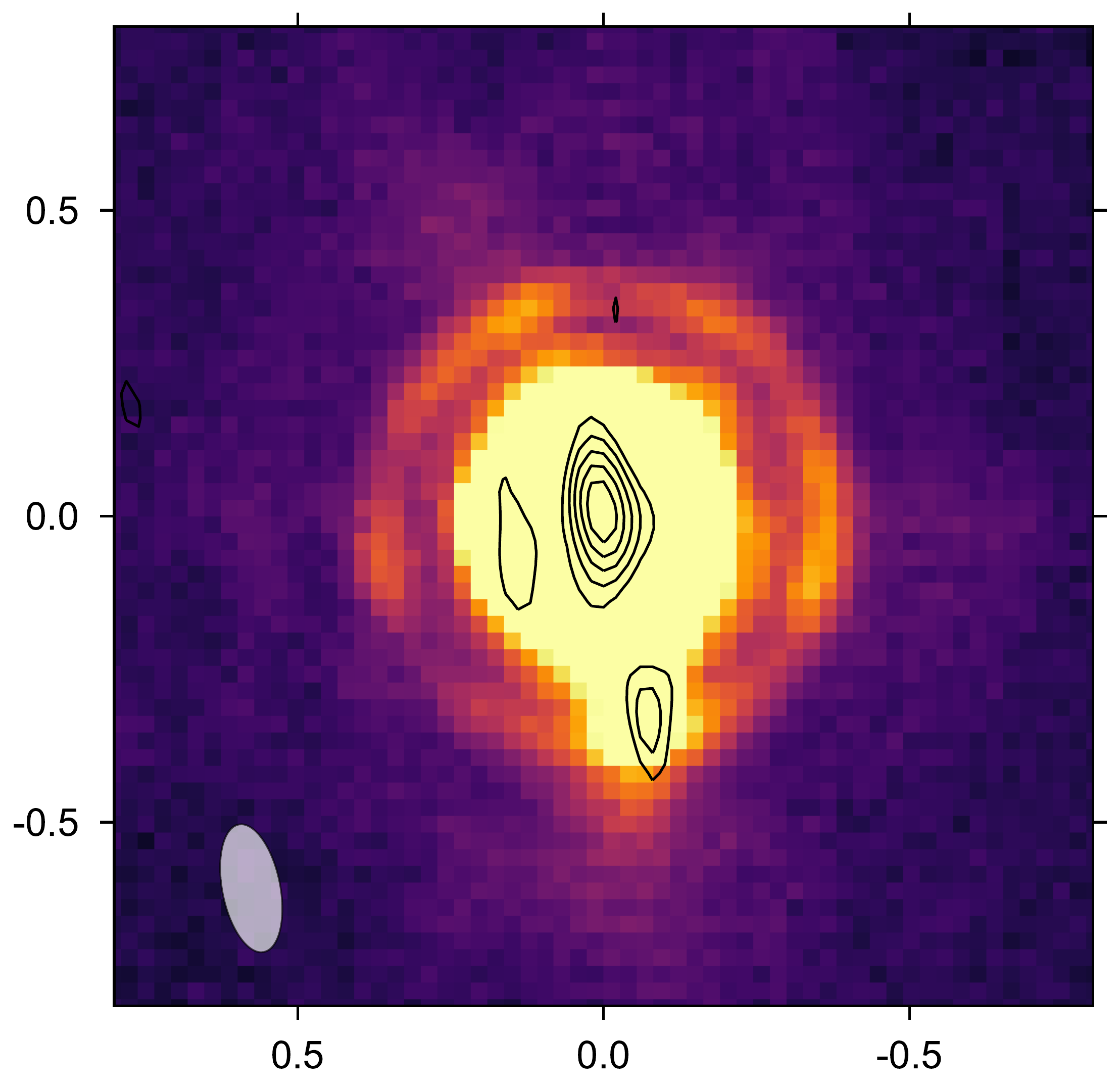}
\caption{$L^\prime$ band reduced imaged. A blob appears in the South-West of the primary, at the location of the gap in the dust. The VLA continuum from \citet{Alves2020} is overplotted with 3$\sigma$ contours levels. The VLA synthesized beam is shown in the bottom left corner. North is up, East is left, scale in arcsec.}
\label{f:pla}
\end{figure}

The $L^\prime$ image, which traces thermal emission as well as scattered light signal, shows a bright peak close to the gap, in the south side of the disk. The blob, which appears in the South-West of the primary in Fig.~\ref{f:pla}, is located at the position of the mm dust gap. To calculate its position we first had to subtract the contribution from the bright primary star (as described in Sec.~\ref{Observations}). { After removing the contribution of the main star it is possible to distinguish the circumstellar disk from the PSF of the companion. Although, some residual light from the circumstellar disk may still be present at the location of the PSF, and the following assumption for its mass is then the upper limit, as its brightness can be overestimated.} The candidate companion is located at a distance of 53 au (0\farcs32), at a position angle of 186$\degree$. This companion, if real, might be responsible for the presence of the gap itself. The flux ratio in $L^\prime$ between the object and the primary star is 1.4$\times$10$^{-2}$, which for the evolutionary models corresponds to an object in a range of { 37--47} M$_{Jup}$, { which is within the mass range estimated by \citet{Alves2020}}. For that assumption we used the range of values of 0.9--5.7 Myr for the age of the system as presented in Sec.~\ref{Stellar_properties}, { a value of 7.93 (derived from the WISE W1 data) for the $L^\prime$ magnitude of the star}, and 166 pc as the distance of the system, which may be slightly different from the real value. The models implied are COND and DUSTY from \citep{Allard2001}, and we also explored the BEX models \citep{2019A&A...623A..85L}.

{ The companion is visible in the $L^\prime$ image only, as in the other filters at shortest wavelengths the contribution of the disk at close separations is higher than the thermal emission of the companion itself (see Table~\ref{Disk_brightness}). For example, in $K_s$ band the same companion would have a contrast of 5.4 mag, and the disk has a comparable brightness at the location of the companion. For each band we checked that the $L^\prime$ detection is consistent with the upper limit of the mass. A similar case is the substellar companion around HD\,169142 presented in \citet{2014ApJ...792L..23R} and \citet{2014ApJ...792L..22B}, detected in the $L^\prime$ band but its signal is not recovered at shortest wavelengths \citep[e.g.,][]{2018MNRAS.473.1774L, 2019A&A...623A.140G}. It is also possible that the  $L^\prime$ emission comes from a circumplanetary disk of a still forming protoplanet. To explore this scenario, H$\alpha$ observations of the source might detect emission from both the primary and the sub-stellar companion.    }

  A comparison with the VLA 14-mm map presented in \citet{Alves2020} shows faint emission at the location of the candidate companion (contours in Fig.~\ref{f:pla}). This 14-mm detection can be non-thermal emission from a brown dwarf companion. The nature of this detection can only be confirmed with a follow-up measurement. Future observations with NIRC2 at Keck, for example, could test whether the object is bound to the system.




\section{Discussion} \label{Discussion}
The results by \citet{Alves2020} and this work portray a peculiar object in the context of planet-forming disks. In fact, the disk of [BHB2007]-1 hosts a prominent cavity (70 au large) that may be explained by the sub-stellar object candidate at 50 au described in Sec.\,\ref{Planet}. To our knowledge, such a wide-separation binary system surrounded by a large circumbinary disk is a very rare configuration although wide orbit giant exoplanets have been reported \citep[e.g.,][]{2010Natur.468.1080M}.

In principle, the large discrepancy between the NIR and mm cavity size ($<$20 au vs 70 au) can be ascribed to the sub-stellar companion. Several models \citep[e.g.,][]{Pinilla2012,2012ApJ...755....6Z} show the differential action exerted by massive companions on micron- and mm-sized dust grains, yielding the former type of grains to filter, where the larger grains are trapped. Comparing our numbers with the modeling by \citet{Villenave2019} reveals that a sub-stellar object ($M>13 {\rm M_{jup}}$) is needed to account for the observed discrepancy. More accurate constraints are deferred to the confirmation and characterization of the companion. 

The presence of a massive companion may also indirectly explain the large brightness difference between the North and South disk sizes (see Sec.\,\ref{Disk}). The constancy of the brightness ratio across wavelengths (see Table \ref{Disk_brightness}) suggests that the difference is due to a different amount of illumination. In other words, the Northern disk side would be partly shadowed while the South side would not. Recently, such a configuration is increasingly observed \citep[see e.g.,][]{Benisty2018, Muro-Arena2020}. These shadows are most likely cast by a misaligned inner disk or disk warp close to the star \citep[as in][]{2015ApJ...798L..44M}. These misalignments would in turn be due to the gravitational force exerted by a companion. A misaligned massive companion, which is more massive than several Jupiter masses, can break the disk and cause the inner and outer disk misalignment \citep[e.g.,][]{2019MNRAS.483.4221Z}. In particular, radiative-transfer models \citep[e.g.,][]{Benisty2018, Nealon2019} reveal that small misalignments are sufficient to cast shadows on very large angles, as is possibly observed around [BHB2007]-1.  Speculatively, the small misalignment between the disk PA \citep[$12.5\degree$ or, equivalently, $192.5\degree$,][]{Alves2020} and the companion PA (186$\degree$, Sec.\,\ref{Planet}) would therefore explain the observed shadowing of the northern disk side. A caveat is that the current distance of the companion may not be the semi-major axis, as what we measure is the {\it projected} separation on the sky. Concerning the cavity size, more robust constraints are entrusted to future observations of this source. As for the NIR infrared, the CO molecular line shows different brightness in between the southern and northern parts.  This is because less illumination means lower temperature so that lines are weak. Other examples of this phenomenon are presented in \citet{10.1093/mnras/stx2523} and \citet{2019MNRAS.483.4221Z}.

An alternative scenario is that the difference in the luminosity is caused by the presence of the binary system [BHB2007]-2. If these systems have formed in the same place and the filament observed is a remnant of the connection that they might have in the past, there might be still gravitational influence in between the two binary systems.

\section{Conclusions}\label{conc}

We present archival VLT/NACO NIR observations of the young system around [BHB2007]-1 and neighboring [BHB2007]-2AB system. The sources are detected in the $J$, $H$, $K_S$, and $L^{\prime}$ broad band filters, and narrow filter NB\_1.64. [BHB2007]-1, part of the Barnard 59 region, most likely shares the same distance of the visual binary system [BHB2007]-2AB. Previous ALMA observations of the system around [BHB2007]-1 show an inclined disk with a large gap fed by extended CO filaments. The NIR data also reveal extended emission in the direction of the southern CO filament, though it is very faint. The [BHB2007]-2AB system appears to be sufficiently close that it externally illuminates the southern filament of [BHB2007]-1. 

[BHB2007]-1's inclined disk is also detected in the NIR and is brighter in the $H$ band with respect to the other filters. The northern and southern extensions of the disk present different surface brightnesses, with the Southern part being $\sim$6 times brighter than the Northern. We speculate that the difference in brightness is due to the presence of a sub-stellar companion. Indeed, $L^{\prime}$ filter images show a close companion with a projected separation of 50 au and a mass in the range { 37--47} M$_{Jup}$ (depending on the age of the system). This emission is coincident with a compact source previously seen with VLA 14-mm observations, and it is located at a separation consistent with being associated with the wide cavity seen in the ALMA mm data. The value of the mass found from the NIR data is compatible with the mass required to open such a large cavity seen with ALMA \citep{Alves2020}. As for the case of the planet PDS\,70b \citep{2018A&A...617A..44K}, these systems provide additional evidence that substellar objects may be the responsible for at least some gaps in protoplanetary disks, but [BHB2007]-1 is unique because the parent star is still in the protostellar phase. This system, along with the detections of gaps in other young protostellar disks, provide intriguing evidence for planet formation occurring far earlier than what is expected in the classical picture of star and planet formation.

\acknowledgments
{ We are very grateful to the referee for their constructive comments and suggestions that significantly improved the quality of this manuscript.}  A.Z. acknowledges support from the FONDECYT Iniciaci\'on en investigaci\'on project number 11190837.
S.P. acknowledges support from the Joint Committee of ESO and the Government of Chile and the FONDECYT Regular grant 1191934. F.O.A. acknowledges financial support from the Max Planck Society.
J.M.G. is supported by the grant AYA2017-84390-C2-R (AEI/FEDER, UE).
G.A.P.F acknowledges partial support from the Brazilian Agency CNPq. L.I.C. acknowledges support from the David and Lucille Packard Foundation, the Virginia Space Grant Consortium, and NASA ATP 80NSSC20K0529.
This paper makes use of the following ESO, ALMA and VLA data: 073.C-0379(A), 081.C-0477(A), DS/JAO.ALMA\#2013.1.00291.S, VLA/16B-290.
ALMA is a partnership of ESO (representing its member states), NSF (USA) and NINS (Japan), together with NRC (Canada), NSC and ASIAA (Taiwan), and KASI (Republic of Korea), in cooperation with the Republic of Chile. The Joint ALMA Observatory is operated by ESO, AUI/NRAO and NAOJ. 
\facilities{VLT, ALMA, VLA}

\bibliography{sample63}{}
\bibliographystyle{aasjournal}

\end{document}